%Paper: cond-mat/9303009
%From: privman@craft.camp.clarkson.edu (Prof. V. Privman)
%Date: Sat, 6 Mar 93 13:47:59 -0500

% The paper is in plain TeX  and it can be
% TeXed directly as provided. No input files,
% etc., are needed.
%
% Two FIGURES are not included to avoid transfer of large Postscript
% files. To get the "figured" addition by hard mail, send an e-mail
% request to PRIVMAN@CRAFT.CAMP.CLARKSON.EDU
%
\def\NI{\noindent}
\def\PPP{\partial}
\long\def\UN#1{$\underline{{\vphantom{\hbox{#1}}}\smash{\hbox{#1}}}$}
\magnification=\magstep 1
\overfullrule=0pt
\hfuzz=16pt
\voffset=0.0 true in
\vsize=8.8 true in
   \def\NP{\vfil\eject}
   \baselineskip 20pt
   \parskip 6pt
   \hoffset=0.1 true in
   \hsize=6.3 true in
\nopagenumbers
\pageno=1
\footline={\hfil -- {\folio} -- \hfil}
\headline={\ifnum\pageno=1 \hfill March 1993 \fi}

\hphantom{AA}

\hphantom{AA}

\centerline{\UN{\bf Crossover from Rate-Equation to
Diffusion-Controlled}}

\centerline{\UN{\bf Kinetics in Two-Particle Coagulation}}

\vskip 0.4in

\centerline{\bf V.~Privman,$^a$\ C.R.~Doering$^a$\ {\rm and}\
H.L.~Frisch$^b$}

\vskip 0.2in

\NI $^a$Department of Physics, Clarkson
University, Potsdam, New York 13699--5820

\NI $^b$Department of Chemistry, State
University of New York, Albany, New York 12222

\vskip 0.6in

\centerline{\bf ABSTRACT}

We develop an analytical diffusion-equation-type approximation
scheme for the one-dimensional coagulation reaction $A+A \to A$
with partial reaction probability on particle encounters which are
otherwise hard-core. The new approximation describes the crossover
from the mean-field rate-equation behavior at short times to the
universal, fluctuation-dominated behavior at large times. The
approximation becomes quantitatively accurate when the
system is initially close to the continuum behavior, i.e.,
for small initial density and fast reaction. For large
initial density and slow reaction, lattice effects are
nonnegligible for an extended initial time interval. In such cases
our approximation provides the correct description of the initial
mean-field as well as the asymptotic large-time,
fluctuation-dominated behavior. However, the intermediate-time
crossover between the two regimes is described only
semiquantitatively.

\vfill

\NI {\bf PACS numbers:}$\;$  82.20.Mj {\it and\/} 05.40.+j

\NP

\NI \UN{\bf 1.~Introduction}

\

Reaction-diffusion systems in low dimensions have been
investigated recently with emphasis on fluctuation-dominated
effects, specifically, the breakdown of the
standard chemical rate equations which correspond to the
``mean-field'' approximation of the reaction kinetics. Much recent
interest has been focused on the simplest reactions of
two-particle coagulation, $A+A \to A$, and annihilation,
$A+A \to {\rm inert}$, on the one-dimensional lattice [1-13].
Indeed, in the diffusion-limited, instantaneous-reaction case, these
processes show non-mean-field power-law decay of the $A$-particle
density in one dimension. Exact results have been derived
indicating that the particle concentration per unit length, $c(t)$,
decays according to

$$ c(t\hbox{-large}) \sim 1/\sqrt{Dt} \; , \eqno(1.1)$$

\NI where $D$ is the single-particle diffusion constant in the
dilute limit, and $t$ is the time. Experimental verification of
the theoretical predictions has been initiated recently [14].

The result (1.1) differs markedly from the prediction of the rate
equation

$$ {dc \over dt} \propto -c^2 \; , \eqno(1.2) $$

\NI appropriate for these reactions. Indeed, the naive
mean-field power law is simply

$$ c_{\hbox{mean-field}}(t\hbox{-large}) \sim \ell /Dt  \; ,
\eqno(1.3) $$

\NI where $\ell$ is the lattice spacing. Modified mean-field treatments
can capture the diffusion-controlled scaling in some cases [15].

Now if the reactions on particle encounters are not instantaneous
but rather occur with probability $q$, where $0 < q \leq 1$, then
for slow reaction, $q \ll 1$, one might expect the mean-field
expressions to apply initially although for large times a
crossover to the fluctuation-dominated behavior (1.1) is expected.
Extensive numerical studies of the annihilation reaction,
$A+A \to {\rm inert}$, with
partial reaction probability [16-17], indicate that there is a
crossover from the mean-field to the fluctuation-dominated
behavior. Similar crossover effects were observed in
related dimer-deposition models with diffusional
relaxation [18]. However, due to lack of any theoretical
description of these crossover effects, the data
interpretation [16-17] has been a fit to an ad-hoc analytical
expression: a $t$-dependent-effective-exponent power law, with no
conclusive results. In fact the analytical form used [16-17] is
known to be wrong in other better understood mean-field to
fluctuation crossovers in static critical phenomena [19], and it is
also inconsistent with a certain dimensional-analysis scaling form
[3] appropriate in the limit of vanishing initial density.

Approximate theoretical description of
the mean-field to fluctuation-dominated crossover is important for
several reasons. Firstly, it will provide guidance on the limits of
validity of the rate-equation approximation and illuminate its
microscopic foundations. Secondly, it will shed light on the
universality of the large-time fluctuation dominated power-law
behavior (1.1). Finally, it can suggest scaling forms and
variable-combinations to use in actual data fits, as well as
directions of improving the simplest mean-field approximation
schemes in those regimes where mean-field rate equations do apply;
see [20] for a discussion of multiparticle reactions for which
data fits can be improved by a careful choice of the
mean-field model variant.

This work reports a new diffusion-equation type approximation
scheme for the description of the mean-field to
fluctuation-dominated crossover in one-dimensional reactions. We
consider the coagulation reaction, $A+A \to A$, rather than
annihilation. Indeed, while these reactions have many common
features [1-13], the former, coagulation reaction, is described
for $q=1$ by a local diffusion equation for the interparticle
distance, in the continuum limit [8]. This provides a convenient
starting point.

When the particles do not react with probability 1, their
interactions upon encounters must be specified. We assume hard-core
interactions in cases of no reaction; our model is defined in
Section 2. It has been anticipated in the literature [21-22] that
a diffusive description of the partial reaction probability system
will involve the so-called ``radiation'' boundary conditions [23].
However, no specific implementations have been reported yielding
results for one-dimensional reactions.

In fact, the form of the
radiation boundary conditions turns out to be the least
straightforward aspect of the problem; the radiation boundary
conditions are discussed in detail in Section 3.
Analytical results within our approximation scheme are derived in
Section 4. Various limiting forms as well as comparison with
numerical Monte Carlo data and concluding discussion are given in
Section 5.

\NP

\NI \UN{\bf 2.~Diffusive Approximation for Interparticle
Distribution}

\

We consider particles hopping on a line, each independently, and
on pairwise encounters subject to reaction $A+A \to A$ with
probability $q$, where $0 \leq q \leq 1$, and to hard-core
interaction (i.e., the hopping attempt fails) with probability
$1-q$. Let $c(t)$ denote particle concentration per unit length at
time $t$, and

$$ c(0) = \rho \eqno(2.1) $$

\NI be the initial concentration. The interparticle distribution
will be denoted by $P(x,t)$. It is proportional to the
probability density to find interparticle distance $x \geq 0$, at
time $t\geq 0$. It will be normalized as follows:

$$ \int\limits_0^\infty P(x,t) dx = { c(t) \over \rho } \; .
\eqno(2.2) $$

\NI We consider a random initial distribution of particles

$$ P(x,0)= \rho e^{-\rho x} \; . \eqno(2.3) $$

Our aim is to develop an approximate diffusion-type description
of the time-evolution of the distribution $P$. Thus we assume that
the time-dependence can be modeled by a probability balance
equation of the form

$$ {\PPP P \over \PPP t} = - {\PPP J \over \PPP x} \; . \eqno(2.4)
$$

\NI The current $J$ will have the diffusive component due to
particle hopping,

$$ J_{\rm diff} = -2D {\PPP P \over \PPP x} \; . \eqno(2.5) $$

\NI Here $D$ is the single-particle diffusion constant in the
dilute limit, i.e., when the hard-core constraint is ineffective in
slowing down the particle hopping. It is well known that the
interparticle distance then evolves with twice the particle
diffusion constant.

In addition, the evolution of $P$ is affected by the reaction,
which essentially enters via the boundary conditions at $x=0$,
and by the hard-core constraint on the particle motion during
those encounters which do not lead to reaction. Let us consider
first the effect of the hard core interactions.

On each such hard-core encounter, illustrated by $x_m = 0$ in
Figure~1, the two interparticle distances nearest to the distance
between the two colliding particles, each cannot evolve purely
diffusively but rather they are driven to shorten. In Figure~1,
the interparticle distances $x_{m-1}$ and $x_{m+1}$ are thus
correlated with each other and with the middle distance $x_m$,
with the correlation being strongest when $x_m$ approaches
zero. While the exact modeling of such correlations might be
quite complicated and involve many-body effects, it can be
modeled approximately, phenomenologically, by assuming a random
(in time and in coordinate $x$) ``driving force'' pushing the
$x$-distances toward the origin of the $x$-axis.

Thus we approximately model the hard-core effects by including a drift
term in the current. The drift velocity is in the negative-$x$
direction, of magnitude conveniently denoted by $v=2D(1-q)p(t)$:

$$ J_{\rm hard\hbox{-}core} = -2D(1-q)p(t)P(x,t) \; . \eqno(2.6) $$

\NI The proportionality to $1-q$ is suggested by the fact that
this factor is the probability of the hard-core interaction
rather than reaction, on encounter. The quantity $p(t)$ was
introduced to represent the fact that the drift effect due to the
hard-core interactions should be proportional to the density of
particle pairs at contact, given by $ P(0,t) $, which suggests that
we put

$$ p(t) = P(0,t) \; . \eqno(2.7) $$

\NI Finally, the dimensionless constant in (2.6)
was determined by the observation that for no reaction (only
hard-core interactions, $q=0$) the stationary solution for all times
must be given by (2.3), while the sum of the diffusive and hard-core
currents, generally given by

$$ J=-2D \left[(1-q)pP+{\PPP P \over \PPP x} \right] \; ,
\eqno(2.8) $$

\NI  must be identically zero for $q=0$.

Let us introduce the reduced time variable

$$ \tau \equiv 2Dt \; . \eqno(2.9) $$

\NI Then we get the equations

$$ {\PPP P \over \PPP \tau} = (1-q)p{\PPP P \over \PPP x} + {\PPP^2
P \over \PPP x^2 } \; , \eqno(2.10) $$

$$ p(\tau )= P(0,\tau ) \; . \eqno(2.11) $$

\NI The first equation follows from (2.4) with (2.8). The solution
is well defined only after the boundary conditions at $x=0$ are
added. The boundary conditions are discussed in Section 3. Note
however that our approximate equation is nonlinear. Thus we are
going to solve it within an adiabatic approximation to be specified
in Section 4.

\NP

\NI \UN{\bf 3.~Radiation Boundary Conditions and Universality}

\

At $x=0$, we assume time-dependent radiation boundary conditions
[23],

$$ J(0,t)=-2Da(t)P(0,t) \; , \qquad\qquad t>0 \; , \eqno(3.1) $$

\NI which correspond to partial absorption of the diffusers. Of
course, in our case the ``absorption'' at the origin means the
disappearance of an interparticle distance due to coagulation
reaction. We will now address the issue of modeling the quantity
$a(t)$ in (3.1).

Generally, the applicability of diffusion-type approximation
schemes improves when the system is close to the continuum limit.
Specifically, if the initial density is small as compared to the
inverse of the microscopic lattice spacing, i.e., $\rho \ell \ll
1$, and the reaction is fast, $(1-q) \ll 1$, then the continuum
approximation should become more accurate. Lattice effects are
more significant for large particle densities, as well as for slow
reactions since the latter are expected to be described by the
mean-field (lattice-dependent) expressions for extended times. For
instantaneous reactions, $q=1$, the hard-core constraint is
irrelevant and the description is in terms of an ordinary, linear
diffusion equation, which then provides an extremely accurate
approximation, exact in the continuum limit [8]. For a discrete-space
modeling of the diffusion-controlled limit, see [24].

Thus we expect that our approximation scheme will also be more
accurate for smaller densities and faster reactions. However, we
cannot fully eliminate the lattice-spacing dependence because our
aim is to describe the crossover from the lattice-dependent
mean-field behavior for short times to the universal,
continuum-limit, fluctuation dominated behavior at large times.

In a lattice hopping-diffuser model, one can show that the quantity
$a$ entering the radiation boundary conditions, is given by

$$ a_{\rm lattice} = {q \over (1-q)\ell } \left[1+O\left({v \ell
  \over D }\right) \right] \; . \eqno(3.2) $$

\NI Here the absorption probability at the origin, $q$, and the
drift velocity towards the origin, $v$, were assumed constant. Thus
initially we can assume this lattice expression in our model as
well,

$$ a(0)=  {q \over (1-q)\ell } \; , \eqno(3.3) $$

\NI where we neglected the higher-order terms in $\ell$.

Quite generally for the coagulation reaction $A+A \to A$, one
anticipates that the density (1.1) approaches a universal
power-law behavior, independent of the initial density and of the
reaction probability $q$,

$$ c(\tau\hbox{-large}) = {1 \over \sqrt{\pi \tau }} \; ,
\eqno(3.4) $$

\NI where the universal coefficient has been calculated [8] for the
case of instantaneous reaction, $q=1$, corresponding to $a=\infty$
in the radiation boundary conditions (3.1). Note that in this
limiting full adsorption case $P(0,t>0) = 0$ and the problem is
just the ordinary diffusion on half-line $x \geq 0$ because there
is no hard-core induced ``drift'' current.

Our numerical Monte Carlo studies, details of which will be
reported in Section 5, support the expectation of the universal
coefficient in (3.4). A simplistic line of argument for this
universality appeals to the recurrence property of the
one-dimensional random walk. Once two particles meet at late times
when the system is dilute, they will have many additional
encounters with each other before being ``mixed'' by diffusion with
other surrounding particles which are distance of order $1/c(t
)$ away. The partial reaction probability is thus expected to be
effectively ``renormalized'' to 1.

Note however that the same
two-particle correlation that leads to these repeated encounters
also perturbs the density locally. In fact, in the problem of two
particles (one interparticle distance) which coagulate
with partial reaction probability, the average particle number,
initially 2, exceeds the final, infinite-time value 1 by an
amount proportional to $1/\sqrt{\tau}$ for large times, with a
nonuniversal coefficient. Thus the universality of the coefficient in
(3.4) is unlikely to be explained at the two-particle level only;
it must be attributed to many-body fluctuation effects. An important
conclusion is that the quantity $a(t)$ in the radiation boundary
conditions must be forced to diverge \UN{\sl phenomenologically}.
It cannot get ``renormalized'' dynamically within the two-particle
relative-distance diffusive model; the $t$-dependence of $a(t)$
must be introduced by hand.

In our calculations we took

$$ a(t)=a(0)\rho /c(t) \; . \eqno(3.5) $$

\NI This choice was favored because it yielded a convenient closed
calculation scheme, to be specified shortly. It is also
dimensionally correct. Furthermore, the effective reaction
probability $q_{\rm effective}(t)$, implied by the choice (3.5),
approaches 1 as $\sim 1/\sqrt{t}$, which seems reasonable in view of
the simplistic two-particle argument presented above.

\NP

\NI \UN{\bf 4.~Solution within the Adiabatic Approximation}

\

The set of relations (2.1), (2.2), (2.3), (2.10), (2.11),
(3.1), (3.5), should be solved for $P(x,t)$ and $c(t)$. In order to
make the calculation tractable, and given the fact that this whole
treatment is not exact but just an approximation scheme, we will
treat  the problem adiabatically. Thus, we first solve (2.10) with
constant $p$, supplemented by the boundary conditions (3.1) with
constant $a$. Given the initial condition (2.3), the resulting
problem is solved for $P(x,t;p,a)$. The relations (2.2), (2.11),
(3.5), evaluated with this ``adiabatic'' $P$-function, provide a
set of three equations for the quantities $p$, $a$, $c$, as
functions of time.

It is convenient to define the quantity

$$ b = (1-q) p \; . \eqno(4.1) $$

\NI It is also useful to recall the relation

$$ a = {q \rho \over (1-q) \ell c} \; , \eqno(4.2) $$

\NI which follows from (3.3) and (3.5). Thus we solve the
diffusion equation

$$ {\PPP P \over \PPP \tau} = b{\PPP P \over \PPP x} + {\PPP^2 P
\over \PPP x^2 } \; , \eqno(4.3) $$

\NI with the initial condition (2.3) and the boundary conditions
at $x=0$ given by

$$ (a-b)P(0,\tau )={\PPP P \over \PPP x}(0,\tau ) \; . \eqno(4.4)
$$

\NI The latter relation follows from (2.8) and (3.1).

The solution $P(x,\tau;a,b)$ is obtained with fixed $a$ and $b$.
Rather cumbersome calculations, not presented in detail here,
yield the Laplace Transform of this function in the
time variable $\tau$. It turns out that the inverse Laplace
Transform of $P(0,\tau;a,b)$ can be obtained in closed form. The
result can then be used to write, by (2.11), the equation

$$ {\rho -a \over \rho } p =
\left(\rho - {b \over 2} \right) e^{\rho (\rho-b) \tau }
{\rm erfc} \left[\left(\rho - {b \over 2} \right) \sqrt{\tau }
\right] - \left(a - {b \over 2} \right) e^{a (a-b) \tau }
{\rm erfc} \left[\left(a - {b \over 2} \right) \sqrt{\tau } \right]
\; . \eqno(4.5) $$

Note that we essentially have two unknowns, $p$ and $c$, since $b$
and $a$ are simply related to $p$ and $c$, respectively. The
second equation connecting these two unknowns follows from (2.2).
In conventional diffusion problems, relation (2.4) for the
probability current can be utilized to reduce the time-derivative
of (2.2) to the form involving flux at the origin,

$$ {dc \over dt} = \rho J(0,t) \; , \eqno(4.6) $$

\NI where $J(0,t)$ can be further replaced by $-2Da(t)P(0,t)$, via
(3.1). The derivation of (4.6) would remain valid for our more
complicated system of relations had we treated it exactly.
However, it is not ensured a priori that the approximate
functions, obtained within our adiabatic scheme,
identically satisfy (4.6). Thus despite its suggestive simplicity,
the relation

$${dc(t) \over d\tau} = -\rho ap \; , \eqno(4.7) $$

\NI which follows from (4.6), was not utilized within our
adiabatic solution scheme. We preferred to use directly the
integral form (2.2) which is presumably more ``robust'' in
preserving physically reasonable properties when setting up
approximations (see further comments in Section 5).

Fortunately, the Laplace Transform of the integral on the
left-hand side of (2.2), calculated within the adiabatic
approximation, can be inverted directly. Again due to complexity
of the intermediate expressions we only quote the final result,
which was further simplified by using (4.5):

$$ {b -a \over \rho } c = p +
{a+\rho-b \over b-\rho} \left\{
\left(\rho - {b \over 2} \right) e^{\rho (\rho-b) \tau }
{\rm erfc} \left[\left(\rho - {b \over 2} \right) \sqrt{\tau }
\right] - {b \over 2}
{\rm erfc} \left( {b \over 2} \sqrt{\tau }
\right) \right\} \; . \eqno(4.8) $$

Relations (4.5) and (4.8) determine the two unknowns $p$ and $c$
as functions of $\tau$. They involve a standard error function,

$$ {\rm erfc}(y)={2 \over \sqrt{\pi} } \int\limits_y^\infty
e^{-z^2} dz \; , \eqno(4.9) $$

\NI and can be solved straightforwardly to numerically obtain the
plot of, for instance, $c(\tau)$, and well as closed-form
analytical expressions in various limits.

\NP

\NI \UN{\bf 5.~Results and Discussion}

\

Numerical Monte Carlo data for the coagulation reaction with
partial reaction probability were obtained for periodic lattices of
size $10^5$. The lattice was randomly populated with the initial
density of particles, $\rho \ell$ per site. Then the particles
hopped randomly and independently, with the rate of the hopping
attempts selected to keep the fixed single-particle (dilute-limit)
diffusion constant, $D$, value. When
a particle attempted to hop to an empty target (nearest neighbor)
site, it was simply moved there. However if the target site
happened to be occupied, the attempting particle was left in the
original site with probability $1-q$, or it was removed
(representing coagulation at the target site) with probability $q$.
The results for the density were recorded and averaged over
$10^3$ independent Monte Carlo runs.

Figure 2 illustrates two typical situations. In the case of
relatively large initial density and slow reaction, i.e., when the
system takes a long time to reach continuum-limiting behavior, the
approximation is semiquantitative at intermediate times although
it is quite accurate for short times, and it reproduces the
correct universal asymptotic behavior at large times. The latter
is not seen in Figure 2, but it can be evaluated analytically; see
below.

In the case of low initial density and fast reaction, the
continuum limiting behavior sets in at earlier times. Our
approximation then follows the data quantitatively, although
there is a small
deviation at intermediate times. In fact such small deviations are
present also in the case of the instantaneous reaction [8] which
is the optimal continuum-limiting situation (see further comments
below) for modeling by the diffusion equation.

For short times, relations (4.5) and (4.8) suggest that the
functions $c(\tau)$ and $p(\tau)$ can be expended in the Taylor
series in $\sqrt{\tau}$. For instance,

$$ p=\rho - {2q\rho \over (1-q)\ell} \left[1+(1-q)\ell \rho
\right] \sqrt{\tau \over \pi} + O(\tau ) \; . \eqno(5.1) $$

\NI The derivation of all but the leading terms in this and other
expansions reported in this section was carried out in
Mathematica, a symbolic computer language. Specifically, the
$O(\tau )$ term in (5.1) has a nonzero, but rather complicated
coefficient.

For the density at short times, we derived the expansion

$$ c=\rho - {q\rho^2 \over (1-q)\ell}\tau +
{4q^2\rho^2 \over 3\sqrt{\pi}(1-q)^2\ell^2} \left[1+(1-q)\ell \rho
\right] \tau^{3/2} + O(\tau^2 ) \; . \eqno(5.2) $$

\NI The absence of the $O\left( \tau^{1/2} \right)$ term in (5.2)
is interesting. Indeed, let us consider the mean-field rate
equation,

$$ {dc \over d\tau } = -\Gamma c^2 \; , \eqno(5.3) $$

\NI where the rate constant is denoted by $\Gamma$. The solution is

$$ c={\rho \over 1+\Gamma \rho \tau}=
\rho - \Gamma \rho^2 \tau + \Gamma^2 \rho^3 \tau^2 +
O\left(\tau^3 \right) \; . \eqno(5.4) $$

\NI Thus the expansion (5.2) suggests that the mean-field behavior
applies ``initially,'' in that the density has the mean-field
slope, corresponding to the rate constant

$$ \Gamma={q \over (1-q) \ell}=a(0) \; , \eqno(5.5) $$

\NI where the latter expression probably applies more generally
than for our particular choice of $a(t)$; cf.~relations (3.3),
(3.5), (4.7). In fact, the form of (4.7) suggests that the
mean-field approximation which amounts to assuming that the
probability of two particles to react is proportional to $c^2$,
in fact takes the form

$$ P(0,t) \simeq P_{\hbox{mean-field}}(0,t)={c^2 (t) \over \rho}
\; . \eqno(5.6) $$

Note that the short-time expansion coefficients diverge as $q \to
1$. In fact our expression for $c(\tau )$ then coincides with the
simple-diffusion results [8] for the instantaneous-reaction case.
The mean-field approximation then breaks down for all $t \geq 0$;
see [8] for details.

We now turn to the large-time behavior. The functions $p(\tau )$
and $c(\tau )$ can be expanded in asymptotic series in powers of
$\tau^{-1/2}$ in this limit. Not surprisingly, the expansion is
ill-defined for $q=0$. Assuming $q>0$, we derived the first two
leading terms,

$$ c(\tau ) \simeq {1 \over \sqrt{\pi \tau} } +
{(1-q) \ell \over \pi q \tau } \; , \eqno(5.7) $$

$$ p(\tau ) \simeq {(1-q) \ell \over 2\pi \rho^2 q \tau^2 } +
 {(1-q)^2 \ell^2 \over \rho^2 q^2 \pi^{3/2} \tau^{5/2} }
 \; . \eqno(5.8) $$

\NI The leading term in $c(\tau )$ is universal as expected.

Finally, both for short times and for large times, we checked that
to order $\tau$ and $1/\tau^2$, in the respective expansions,
the relation (4.7) in not violated by the adiabatic-approximation
results, i.e., the difference of the left- and right-hand sides is
of order higher than the powers indicated. However, no careful
numerical or analytical tests were attempted at intermediate times.

In summary we proposed an analytical approximation scheme to
describe the crossover from the initially mean-field to
asymptotically fluctuation-dominated behavior in one-dimensional
two-particle coagulation reactions. The resulting expressions are
not simple. However, they suggest the form of the small- and
large-time behaviors, i.e., the appropriate variable combinations
which are presumably quite general. They also indicate
that any ``scaling-form'' description of the crossover must allow
for at least two scaling combinations,

$$ \rho \sqrt{Dt} \qquad {\rm and} \qquad
{q\sqrt{Dt} \over (1-q)\ell} \; . \eqno(5.9) $$

\NI The approximation scheme used in this work is not an obvious
``leading order'' in a systematic expansion but rather a
phenomenological closed-form model. It is hoped that this first
theoretical study will stimulate further work on
partial-reaction-probability diffusion-reaction systems and
generally on the mean-field to fluctuation-dominated crossover in
nonequilibrium dynamical models.

This research has been supported in part by the NSF grants
PHY-8958506, DMR-9023541 and PHY-9214715.

\NP

\centerline{\bf REFERENCES}

{\frenchspacing

\item{[1]} M. Bramson and D. Griffeath,
Ann. Prob. {\bf 8}, 183 (1980).

\item{[2]} D.C. Torney and H.M. McConnell,
J. Phys. Chem. {\bf 87}, 1941 (1983).

\item{[3]} K. Kang, P. Meakin, J.H. Oh and S. Redner,
J. Phys. A{\bf 17}, L665 (1984).

\item{[4]} T. Liggett, {\sl Interacting Particle Systems\/}
(Springer-Verlag, New York, 1985).

\item{[5]} Z. Racz, Phys. Rev. Lett. {\bf 55}, 1707 (1985).

\item{[6]} A.A. Lushnikov, Phys. Lett. A{\bf 120}, 135 (1987).

\item{[7]} M. Bramson and J.L. Lebowitz, Phys. Rev. Lett. {\bf 61},
2397 (1988).

\item{[8]} C.R. Doering and D. ben--Avraham, Phys. Rev. A{\bf 38},
3035 (1988).

\item{[9]} V. Kuzovkov and E. Kotomin, Rep. Prog. Phys.
{\bf 51}, 1479 (1988).

\item{[10]} J.L. Spouge, Phys. Rev. Lett. {\bf 60},
871 (1988).

\item{[11]} J.G. Amar and F. Family, Phys. Rev. A{\bf 41}, 3258
(1990).

\item{[12]} D. ben--Avraham, M.A. Burschka and C.R. Doering,
J. Stat. Phys. {\bf 60}, 695 (1990).

\item{[13]} V. Privman, J. Stat. Phys. {\bf 69}, 629 (1992).

\item{[14]} R. Kroon, H. Fleurent and R. Sprik, {\sl
Diffusion-Limited Exciton Fusion Reaction in One-Dimensional
Tetramethylammonium Manganese Trichloride (TMMC)}, Phys. Rev. E (in
print).

\item{[15]} J.C. Lin, C.R. Doering and D. ben--Avraham,
Chem. Phys. {\bf 146}, 355 (1990).

\item{[16]} L. Braunstein, H.O. Martin, M.D. Grynberg
and H.E. Roman, J. Phys. A{\bf 25}, L255 (1992).

\item{[17]} H.O. Martin and L. Braunstein, {\sl Study of $A+A \to
0$ with Probability of Reaction and Diffusion in One Dimension and
in Fractal Substrata}, preprint.

\item{[18]} V.~Privman and P.~Nielaba, Europhys. Lett. {\bf 18},
673 (1992).

\item{[19]} P. Seglar and M.E. Fisher, J. Phys. C{\bf 13}, 6613
(1980).

\item{[20]} V. Privman and M.D. Grynberg, J. Phys. A{\bf 25}, 6575
(1992).

\item{[21]} H. Teitelbaum, S. Havlin and G. Weiss, Chem. Phys. {\bf
146}, 351 (1990).

\item{[22]} H. Teitelbaum, R. Kopelman, G. Weiss and S. Havlin,
Phys. Rev. A{\bf 41}, 3116 (1990).

\item{[23]} F.C. Collins and G.E. Kimball,
J. Colloid Sci. {\bf 4}, 425 (1949).

\item{[24]} J.C. Lin, Phys. Rev. A{\bf 45}, 3892 (1992).

}

\NP

\centerline{\bf FIGURE CAPTIONS}

\

\noindent\hang{\bf Fig.~1.}\ \ A schematic representation of the
diffusive motion of four particles. When two middle particles run
into each other but do not react, their interparticle spacing,
$x_m$, bounces off zero, while the nearby interparticle distances,
$x_{m\pm 1}$, receive a ``push'' towards zero.

\

\noindent\hang{\bf Fig.~2.}\ \ Numerical Monte Carlo results,
solid lines, compared to the approximate density $c(\tau)$
calculated from (4.5) and (4.8), dashed lines. The upper curves
correspond to a typical case of initially strong lattice effects,
$\rho=0.8$ and $q=0.1$ (large initial density and slow reaction),
while the lower curves correspond to the case of near-continuum
initial conditions, $\rho=0.1$ and $q=0.9$, i.e., low density and
fast reaction. The latter data were multiplied by 2 for better
resolution. The densities are in units of $1/\ell$, while $\tau$ is
measured in units of $\ell^2$.

\bye